\documentclass[
twocolumn,
showpacs,
preprintnumbers,
amsmath,
amssymb
]{revtex4-1}

\usepackage{graphicx,latexsym}
\usepackage{amsmath,amssymb,amsfonts}
\usepackage{bm}
\usepackage{braket}
\usepackage{mathrsfs}
\newcommand{\up}{\uparrow}
\newcommand{\down}{\downarrow}
\newcommand{\m}{\mathcal}
\newcommand{\ep}{\varepsilon}
\begin{document}
\title{Engineering nonequilibrium superconducting phases in a voltage-driven superconductor under an external magnetic field}
\author{Taira Kawamura\email{tairakawa@g.ecc.u-tokyo.ac.jp}$^{1}$ and Yoji Ohashi$^{2}$}
\affiliation{$^1$ Department of Basic Science, The University of Tokyo, 3-8-1 Komaba, Tokyo 153-8902, Japan}
\affiliation{$^2$Department of Physics, Keio University, 3-14-1 Hiyoshi, Kohoku-ku, Yokohama 223-8522, Japan}
\date{\today}

\begin{abstract}

We theoretically investigate nonequilibrium properties of a normal metal-superconductor-normal metal (NSN) junction under an external magnetic field. When a bias voltage is applied between the normal-metal leads, the confined superconductor is driven out of equilibrium, resulting in a nonequilibrium quasiparticle distribution function having a two-step structure. Using the nonequilibrium Green’s function technique, we determine a comprehensive phase diagram of the nonequilibrium superconductor. Our analysis reveals that the interplay between Zeeman-split energy bands and the nonequilibrium distribution function gives rise to a rich phase structure. Notably, we find that superconductivity destroyed by a strong external magnetic field revives by applying the bias voltage. This reentrant phenomenon is shown to originate from four effective ``Fermi surfaces" that result from the combination of Zeeman-split energy bands and the two-step structure in the nonequilibrium distribution function. Our results demonstrate the possibility of controlling quantum states of matter through the combined engineering of energy band structures and distribution functions.

\end{abstract}

\maketitle
\section{Introduction}

\begin{figure}[t]
\centering
\includegraphics[width=7.8cm]{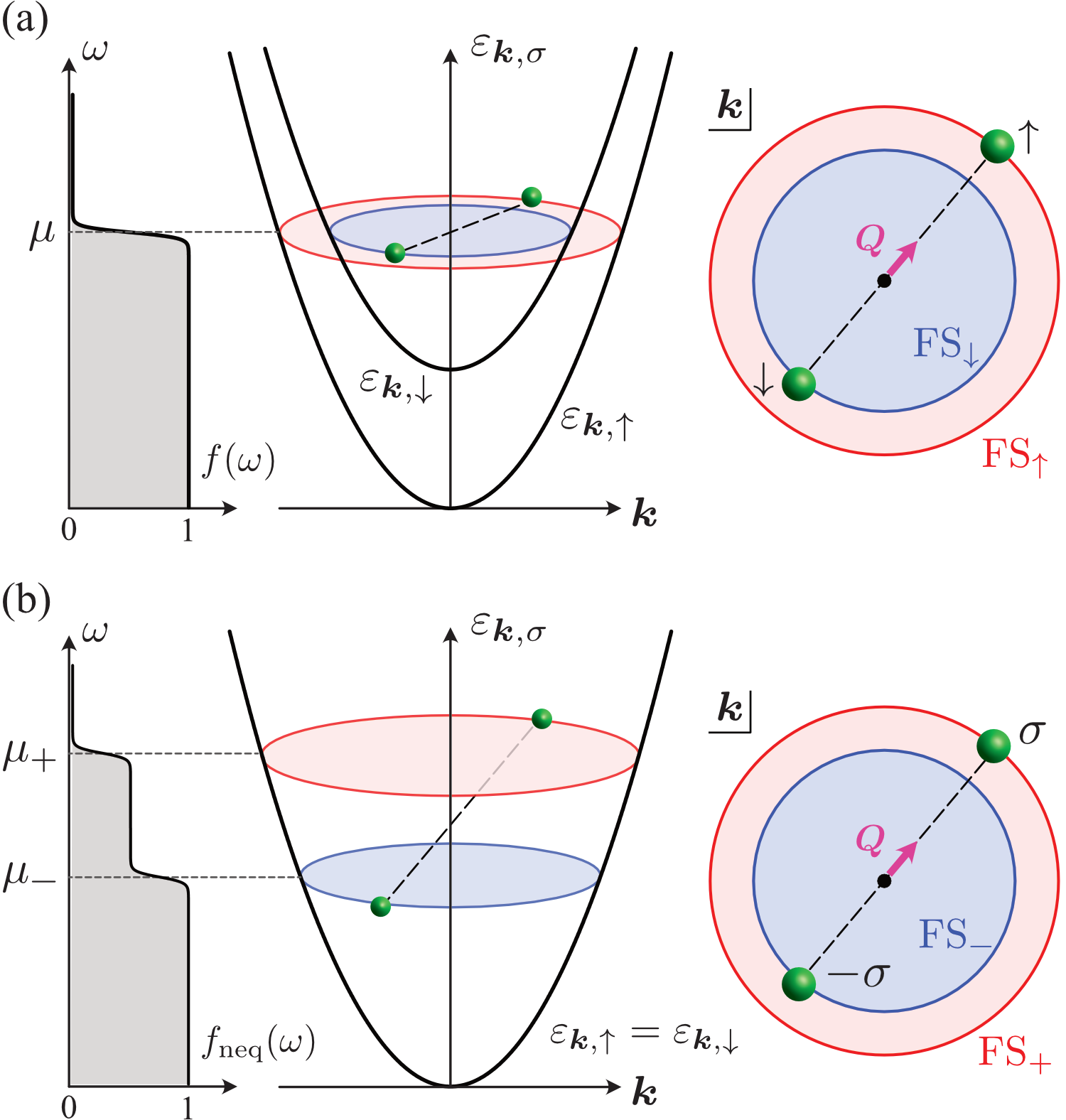}
\caption{Mechanisms for Cooper pairs with nonzero center-of-mass momentum $\bm{Q}$. (a) Conventional FFLO state induced by Zeeman-split energy bands $\ep_{\bm{k},\up} \neq \ep_{\bm{k},\down}$. (b) FFLO-type inhomogeneous superconducting state induced by the two-step nonequilibrium distribution function $f_{\rm neq}(\omega)$. While the energy bands are degenerate ($\ep_{\bm{k},\up}=\ep_{\bm{k},\down}$) in the absence of a Zeeman field, the two edges in the nonequilibrium distribution function $f_{\rm neq}(\omega)$ at $\omega = \mu_\pm$ work as two effective ``Fermi surfaces" ($\textrm{FS}_+$ and $\textrm{FS}_-$) with different sizes.
}
\label{fig.mechanism} 
\end{figure}

Understanding unconventional superconductivity is a major challenge in condensed matter physics~\cite{Sigrist1991, Mineev1999, Scalapino2012}. Among various pairing mechanisms beyond the Bardeen-Cooper-Schrieffer (BCS) theory, the Fulde-Ferrell-Larkin-Ovchinnikov (FFLO) state is notable for its unique feature of Cooper pairs with nonzero center-of-mass momentum $\bm{Q}$~\cite{Fulde1964, Larkin1964, Takada1969, Matsuda2007}. Under a strong external magnetic field, the energy bands $\ep_{\bm{k}, \sigma}$ of electrons with spin $\sigma=\uparrow, \downarrow$ split due to the Zeeman effect, as illustrated in Fig.~\ref{fig.mechanism}(a). The mismatch between the Fermi surfaces resulting from these Zeeman-split energy bands induces Cooper pairs with momentum $\bm{Q}$, leading to a spatially modulated FFLO superconducting order parameter $\Delta \sin(\bm{Q}\cdot \bm{r})$. Experimental signatures of the FFLO state have recently been reported in various materials, including heavy fermion compounds $\textrm{CeCoIn}_5$~\cite{Bianchi2002, Bianchi2003, Kumagai2006, Lin2020, Kittaka2023} and $\textrm{CeCu}_2\textrm{Si}_2$~\cite{Kitagawa2018}, organic superconductors~\cite{Singleton2000, Wright2011, Mayaffre2014, Sugiura2019, Imajo2022, Molatta2024}, as well as iron-based superconductors $\textrm{KFe}_2\textrm{As}_2$~\cite{Cho2017} and $\textrm{FeSe}$~\cite{Ok2020, Kasahara2020, Kasahara2021}. Moreover, the possibility of FFLO-type inhomogeneous Fermi superfluids has also been extensively explored in spin-imbalanced ultracold Fermi gases ~\cite{Hu2006, Parish2007, Sheehy2007, Gubbels2013, Baarsma2013, Kinnunen2018, Kawamura2022, Liao2010}.

In recent years, alternative mechanisms for realizing FFLO-type inhomogeneous superconducting states without relying on Zeeman-split energy bands have garnered significant attention~\cite{Doh2006, Vorontsov2006, He2018, Zheng2015, Zheng2016, Nocera2017, Sumita2016, Sumita2017, Sumita2023, Amin2024, Moor2009, Bobkova2013, Kawamura2022_Jul, Kawamura2024, Kawamura_AAPPS}. One such mechanism involves a nonequilibrium distribution function~\cite{Moor2009, Bobkova2013, Kawamura2022_Jul, Kawamura2024, Kawamura_AAPPS}. In a nonequilibrium fermion system confined between two reservoirs with different chemical potentials ($\mu_+$ and $\mu_-$), fermions follow a nonequilibrium distribution function $f_{\rm neq}(\omega)$, which deviates from the equilibrium Fermi-Dirac distribution function $f(\omega)=[1+e^{\omega/T}]^{-1}$. In particular, as illustrated in Fig.~\ref{fig.mechanism}(b), at low temperatures, $f_{\rm neq}(\omega)$ exhibits a two-step structure that reflects the different chemical potentials $\mu_\pm$ in the two reservoirs. This ``two-step distribution function" has been experimentally observed in a voltage-biased metal wire sandwiched between two electrodes with different electrochemical potentials~\cite{Pothier1997, Franceschi2002, Tikhonov2020}. Similar two-step structures have also been reported in voltage-biased carbon nanotubes~\cite{Chen2009, Bronn2013}. Moreover, their potential realization in ultracold Fermi gases in a two-terminal configuration has also been explored~\cite{Lebrat2018, Mohan2024}.

Multi-step structures in nonequilibrium distribution functions are known to work like multiple effective ``Fermi surfaces"~\cite{Kawamura2022_Jul, Kawamura2024, Kawamura_AAPPS, Kawamura2020, Abanin2005, Shi2024, Ono2019}. These effective ``Fermi surfaces" give rise to various unconventional phenomena, including anomalous Fermi-edge singularity~\cite{Abanin2005}, unconventional quantum oscillation~\cite{Shi2024}, and enhancement of susceptibility at the nesting vector of ``Fermi surfaces"~\cite{Ono2019}. The FFLO-type inhomogeneous superconducting state induced by the two-step distribution function can also be understood from the viewpoint of the existence of these effective ``Fermi surfaces". As illustrated in Fig.~\ref{fig.mechanism}(b), even in the absence of a Zeeman field (where the spin $\sigma=\uparrow, \downarrow$ energy bands remain degenerate), effective ``Fermi surfaces" ($\textrm{FS}_+$ and $\textrm{FS}_-$) with different sizes emerge at the edges $\omega=\mu_\pm$ in the nonequilibrium two-step distribution function $f_{\rm neq}(\omega)$. Cooper pairs formed between FS$_+$ and FS$_-$ acquire nonzero center-of-mass momentum $\bm{Q}$, leading to an FFLO-type inhomogeneous superconducting state. In the conventional FFLO state, the Cooper pair momentum $|\bm{Q}|$ is dominated by the magnitude of the Zeeman field~\cite{Fulde1964, Larkin1964, Takada1969, Matsuda2007}. In contrast, in the nonequilibrium case, $|\bm{Q}|$ depends on the potential difference $\mu_+ - \mu_-$ between the two reservoirs~\cite{Kawamura2024}.

\begin{figure}[t]
\centering
\includegraphics[width=7.8cm]{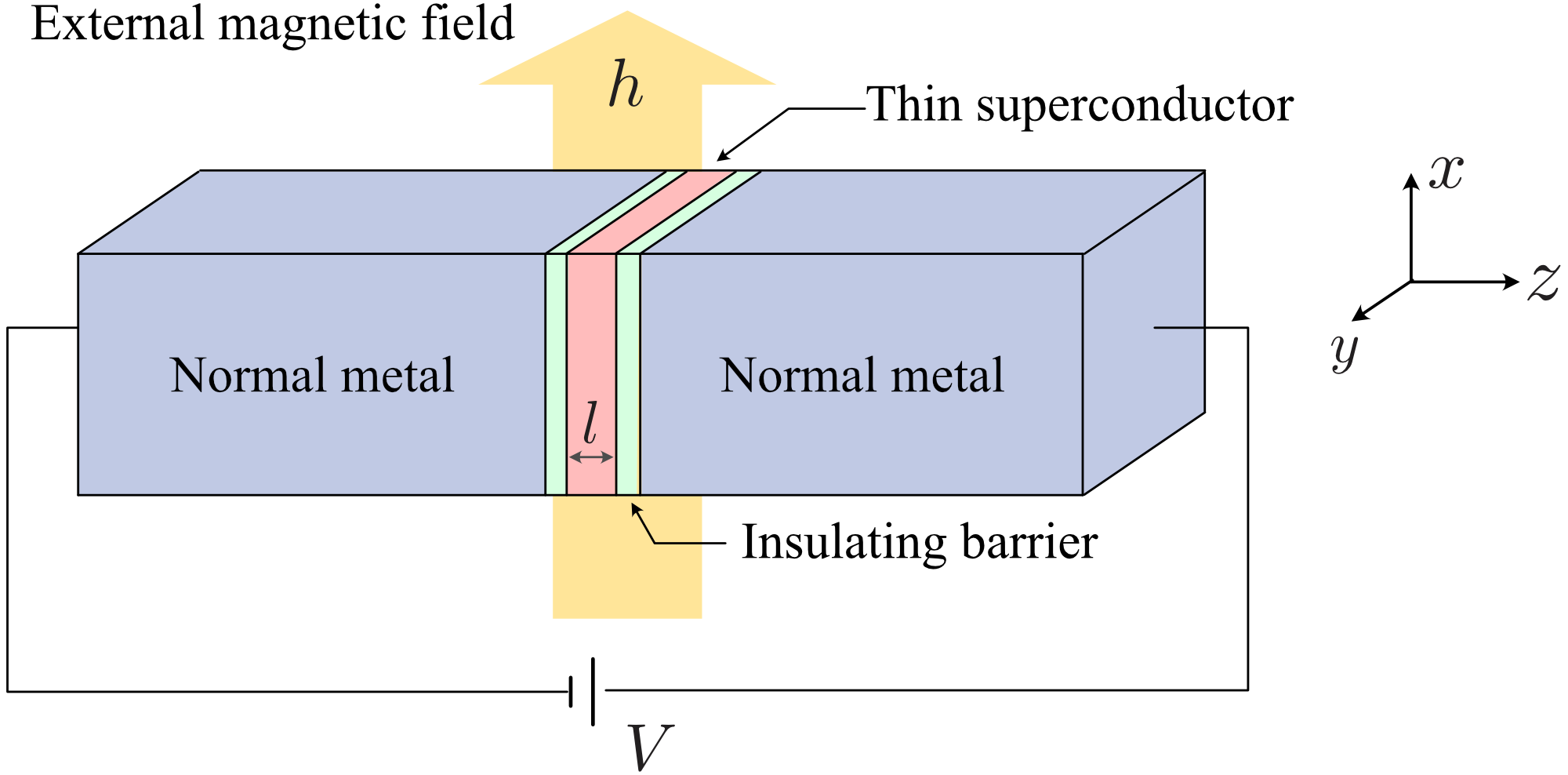}
\caption{Schematic of the device under consideration. A thin ($l \lesssim \xi$) superconductor is confined between left and right normal-metal leads, which can be approximated as isolated systems in thermal equilibrium. The superconductor is driven out of equilibrium by applying a bias voltage $V$ across the leads. An external magnetic field $h$ is applied parallel to the $x$-$y$ plane of the device.}
\label{fig.model} 
\end{figure}

In this paper, we investigate the interplay between Zeeman-split energy bands $\ep_{\bm{k},\up}\neq \ep_{\bm{k}, \down}$ and the two-step distribution function $f_{\rm neq}(\omega)$ in a nonequilibrium superconductor under an external magnetic field. To this end, we consider a model normal metal–superconductor–normal metal (NSN) junction, where a thin ($l \lesssim \xi$) superconductor is confined between two thick normal-metal leads, as illustrated in Fig.~\ref{fig.model}. The superconductor is driven out of equilibrium by the bias voltage $V$ applied across the two normal-metal leads. An external magnetic field $h$, applied parallel to the $x$-$y$ plane of the device, induces Zeeman splitting in the energy bands while minimizing orbital pair-breaking effects~\cite{Matsuda2007, Shimahara1997, Shimahara2009}. Using the nonequilibrium Green's function technique~\cite{RammerBook, ZagoskinBook, StefanucciBook}, we determine the comprehensive phase diagram of the superconductor under these conditions. Our results reveal that the interplay between Zeeman-split energy bands and the two-step distribution function gives rise to a rich phase structure. In particular, we find that superconductivity suppressed by a strong external magnetic field revives by driving the system out of equilibrium. We point out that this superconducting reentrant phenomenon is due to the combined effect of the mechanisms illustrated in Figs.~\ref{fig.mechanism}(a) with (b).

We note that several previous studies have investigated models similar to the one shown in Fig.~\ref{fig.model}~\cite{Kawamura2023, Ouassou2018}. However, Ref.~\cite{Kawamura2023} focuses solely on the superconducting phase transition temperature $T_{\rm c}$ without examining the superconducting phase below $T_{\rm c}$. In Ref.~\cite{Ouassou2018}, the possibility of the FFLO-type inhomogeneous superconducting state is not considered. Moreover, neither study provides a comprehensive stability analysis of nonequilibrium steady states (NESSs)~\cite{Kawamura2023, Ouassou2018}. In this work, we address these limitations by extending the theoretical frameworks developed in Ref.~\cite{Kawamura2024} to include external magnetic field effects.

This paper is organized as follows. In Sec.~\ref{sec.formalism}, we explain our formalism based on the nonequilibrium Green's function technique. We then draw the nonequilibrium superconducting phase diagram in Sec.~\ref{sec.result}. Throughout this paper, we use units such that $\hbar= k_{\rm B}= 1$, and the volumes of the normal-metal leads are taken to be unity, for simplicity.

\section{Formalism \label{sec.formalism}}

\subsection{Model Hamiltonian}

We consider a NSN junction subject to an external magnetic field $h$, as depicted in Fig.~\ref{fig.model}. To model this device, we consider the Hamiltonian,
\begin{equation}
H=H_{\rm SC} + H_{\rm lead} +H_{\rm mix}.
\label{eq.Htot}
\end{equation}
Here, $H_{\rm SC}$ describes the superconductor (main system)  confined between two normal-metal leads. We assume that the superconductor is thin ($l\lesssim \xi$), so that the spatial variation of the superconducting order parameter along the $z$ direction can be ignored. Under this assumption, the superconductor can be modeled by the two-dimensional attractive Hubbard model on a square lattice of $N=L_x \times L_y$ sites with the periodic boundary conditions, given by
\begin{align}
H_{\rm SC} &= H_0 +H_{\rm int} \label{eq.Hsc}
,\\
H_0 &=
-t_\parallel \sum_{\sigma=\up, \down} \sum_{(j,k)}\big[c^\dagger_{j,\sigma} c_{k,\sigma} +{\rm H.c.} \big] 
\notag\\
&\hspace{2cm}
-\sum_{\sigma=\up,\down}\sum_{j=1}^{N}\big[\mu_{\rm sys}+\sigma h\big] n_{j,\sigma}  
\notag\\
&= \hat{\bm{\Psi}}^\dagger \hat{\bm{\m{H}}}_0\hat{\bm{\Psi}}
\label{eq.H0}
,\\
H_{\rm int} &= -U \sum_{j=1}^N n_{j,\up} n_{j,\down}.
\label{eq.Hint}
\end{align}
Here, $c_{j,\sigma}$ is the annihilation operator of an electron with spin $\sigma=\up, \down$ at site $j$ ($=1,\cdots,N$), $n_{j,\sigma}=c^\dagger_{j,\sigma} c_{j,\sigma}$ is the number operator, and 
\begin{equation}
\hat{\bm{\Psi}}^\dagger = \big(\bm{\Psi}_1, \cdots, \bm{\Psi}_N \big)
\end{equation}
with $\bm{\Psi}^\dagger_j=\big(c^\dagger_{j,\up}, c_{j,\down}\big)$ is the $2N$-component Nambu field. In Eq.~\eqref{eq.H0}, $-t_\parallel$ represents the nearest-neighbor hopping amplitude and the summation $(j,k)$ is taken over the nearest-neighbor sites. The parameter $\mu_{\rm sys}$ in Eq.~\eqref{eq.H0} tunes the filling fraction of the main system, and $-U$ ($<0$) in Eq.~\eqref{eq.Hint} characterizes the strength of a Hubbard-type on-site pairing interaction.

The left ($\alpha$=L) and right ($\alpha$=R) normal-metal leads are described by $H_{\rm lead}$ in Eq.~\eqref{eq.Htot}, having the form,
\begin{equation}
H_{\rm lead} = \sum_{\alpha={\rm L}, {\rm R}} \sum_{\sigma=\up, \down} \sum_{j=1}^N \xi^\alpha_{\bm{p},\sigma}(t) d^{\alpha\dagger}_{j, \bm{p}, \sigma} d^{\alpha}_{j, \bm{p}, \sigma},
\end{equation}
where $d^\alpha_{j, \bm{p}, \sigma}$ creates an electron with the kinetic energy $\xi^\alpha_{\bm{p},\sigma}(t)=\bm{p}^2/(2m) -\mu_{\alpha, \sigma}(t)$ in the $j$th $\alpha$ reservoir. We model the coupling between the superconductor and the $\alpha$ normal-metal lead by assuming each site of the main system is coupled to an independent free-fermion $\alpha$ reservoir. These reservoirs are assumed to be in the thermal equilibrium state characterized by electrochemical potential $\mu_{\alpha, \sigma}(t)$ and temperature $T_{\rm env}$. The potential difference $\mu_{{\rm L}, \sigma}(t)-\mu_{{\rm R},\sigma}(t)$ equals the applied bias voltage $eV(t)$ imposed between the two normal-metal leads. For later convenience, we parametrize $\mu_{\alpha, \sigma}(t)$ as
\begin{align}
\mu_{{\rm L},\sigma}(t) 
&= 
\mu_{\rm res} +\sigma h + \frac{e}{2}V(t)
\notag\\
&=
\mu_{\rm res} +\sigma h + \frac{e}{2}
\big[V_0 +\Theta(t)\Delta V(t) \big],\\
\mu_{{\rm R},\sigma}(t) 
&= 
\mu_{\rm res} +\sigma h - \frac{e}{2}
\big[V_0 +\Theta(t)\Delta V(t) \big],
\end{align}
where $\Theta(t)$ is the step function. We assume that the normal-metal leads are connected to the superconductor at $t=-\infty$, and the system is in a NESS under a constant voltage $V_0$ at $t=0$. For $t>0$, the superconductor is driven by a time-dependent voltage $V(t)=V_0+\Delta V(t)$.

The couplings between the superconductor and the normal-metal leads are described by
\begin{equation}
H_{\rm mix} = -\sum_{\alpha={\rm L}, {\rm R}} \sum_{\sigma=\up, \down} \sum_{j=1}^N \sum_{\bm{p}} \big[t_{\perp,\sigma}^{\alpha} d^{\alpha\dagger}_{j, \bm{p}, \sigma} c_{j, \sigma} +{\rm H.c.} \big],
\end{equation}
where $-t_{\perp, \sigma}^\alpha$ is the hopping amplitude between the superconductor and the $\alpha$ reservoir. In this paper, for simplicity, we ignore the spin dependence ($t_{\perp, \sigma}^\alpha =t_{\perp}^\alpha$) and consider the symmetric hopping amplitude $t_{\perp}^{\rm L}=t_{\perp}^{\rm R} \equiv t_\perp$.

\subsection{Nonequilibrium Green's function}
To determine the phase diagram of the nonequilibrium superconductor, we employ the nonequilibrium Green's function tcheqnique~\cite{RammerBook, ZagoskinBook, StefanucciBook}. We introduce $2N\times 2N$ matrix nonequilibrium Green's functions, given by
\begin{align}
\hat{\bm{G}}^{\m{R}}(t,t')
&=
\big[\hat{\bm{G}}^{\m{A}}(t',t) \big]^\dagger
\label{eq.GRA}
\notag\\
&=
-i\Theta(t-t')
\braket{\big[\hat{\bm{\Psi}}(t), \hat{\bm{\Psi}}^\dagger(t')\big]_+}
,\\
\hat{\bm{G}}^<(t,t')
&=
i \braket{\hat{\bm{\Psi}}^\dagger(t') \hat{\bm{\Psi}}(t)},
\label{eq.G<}
\end{align}
where $[A, B]_\pm = AB \pm BA$. In Eqs.~\eqref{eq.GRA} and \eqref{eq.G<}, $\hat{\bm{G}}^{\m{R}}$, $\hat{\bm{G}}^{\m{A}}$, and $\hat{\bm{G}}^<$ represent the retarded, advanced, and lesser Green's functions, respectively. These are computed by solving the Dyson-Keldysh equations~\cite{RammerBook, ZagoskinBook, StefanucciBook}, given by
\begin{align}
&
\hat{\bm{G}}^{\m{R}(\m{A})}(t,t')=
\hat{\bm{G}}^{\m{R}(\m{A})}_0(t,t')
+
\int_{-\infty}^\infty dt_1
\int_{-\infty}^\infty dt_2\hspace{0.1cm}
\notag\\
&\hspace{1cm}\times
\hat{\bm{G}}^{\m{R}(\m{A})}_0(t,t_1)
\hat{\bm{\Sigma}}^{\m{R}(\m{A})}(t_1,t_2)
\hat{\bm{G}}^{\m{R}(\m{A})}(t_2,t')
\label{eq.Dyson.eq}
,\\[4pt]
&
\hat{\bm{G}}^<(t,t')=
\int_{-\infty}^\infty dt_1
\int_{-\infty}^\infty dt_2\hspace{0.1cm}
\hat{\bm{G}}^{\m{R}}(t,t_1)
\hat{\bm{\Sigma}}^<(t_1,t_2)
\notag\\
&\hspace{1cm}\times
\hat{\bm{G}}^{\m{A}}(t_2,t')
\label{eq.Keldysh.eq}.
\end{align}
Here, 
\begin{equation}
\hat{\bm{G}}^{\m{R}(\m{A})}_0 (t, t')=
\int_{-\infty}^\infty \frac{d\omega}{2\pi} e^{-i\omega(t-t')}
\frac{1}{\omega_\pm -\hat{\bm{\m{H}}}_0}
\label{eq.Gr0}
\end{equation}
is the Green’s function describing the isolated superconductor in the thermal equilibrium state before the system-lead couplings, as well as the pairing interaction, are switched on. In Eq.~\eqref{eq.Gr0}, $\omega_\pm = \omega \pm i\delta$, where $\delta$ is an infinitesimally small positive number, and $\hat{\bm{\m{H}}}_0$ is given in Eq.~\eqref{eq.H0}.

The self-energy $\hat{\bm{\Sigma}}^{{\rm X}=\m{R}, \m{A}, <}$ appearing in Eqs.~\eqref{eq.Dyson.eq} and \eqref{eq.Keldysh.eq} consists of two contributions,
\begin{equation}
\hat{\bm{\Sigma}}^{\rm X}(t,t')=
\hat{\bm{\Sigma}}^{\rm X}_{{\rm int}}(t,t')+
\hat{\bm{\Sigma}}^{\rm X}_{{\rm lead}}(t,t').
\label{eq.self}
\end{equation}
The first term $\hat{\bm{\Sigma}}^{\rm X}_{{\rm int}}$ accounts for effects of the pairing interaction, while the second term $\hat{\bm{\Sigma}}^{\rm X}_{{\rm lead}}$ describes effects of couplings to the normal-metal leads. In this work, we evaluate $\hat{\bm{\Sigma}}^{\rm X}_{{\rm int}}$ and $\hat{\bm{\Sigma}}^{\rm X}_{{\rm lead}}$ within the mean-field BCS approximation and the second-order Born approximation with respect to the hopping amplitude $t_\perp$, respectively. In these approximations, the self-energies have the forms~\cite{Kawamura2024},
\begin{align}
\hat{\bm{\Sigma}}^{\m{R}}_{\rm int}(t,t') 
&=
\hat{\bm{\Sigma}}^{\m{A}}_{\rm int}(t,t')
= -\hat{\bm{\Delta}}(t)\delta(t-t')
\label{eq.self.int.R}
,\\
\hat{\bm{\Sigma}}^{<}_{{\rm int}}(t,t') &= 0
,\\
\hat{\bm{\Sigma}}^{\m{R}}_{\rm lead}(t,t') 
&=
\big[\hat{\bm{\Sigma}}^{\m{A}}_{\rm lead}(t',t) \big]^\dagger
=
-2i\gamma \delta(t-t') \hat{\bm{1}}
\label{eq.self.lead.R}
,\\
\hat{\bm{\Sigma}}^<_{\rm lead}(t,t') 
&=
2i\gamma \sum_{\eta=\pm} \exp\left(-i\eta \int_{t'}^t dt_1 \frac{e\Delta V(t_1)}{2}\right) 
\notag\\
&\hspace{1cm}\times
\int_{-\infty}^\infty \frac{d\omega}{2\pi} e^{-i\omega(t-t')} f\left(\omega -\eta \frac{eV_0}{2}\right)\hat{\bm{1}}.
\label{eq.self.lead.<}
\end{align}
For a detailed derivation of these self-energies, see Ref.~\cite{Kawamura2024}. In Eq.~\eqref{eq.self.int.R}, 
\begin{equation}
\hat{\bm{\Delta}}(t)=
\begin{pmatrix}
\bm{\Delta}_1(t) &  \\
 & \ddots \\
&& \bm{\Delta}_N(t)
\end{pmatrix}
\end{equation}
is a $2N\times 2N$ matrix superconducting order parameter, where
\begin{align}
\bm{\Delta}_j(t) 
&=
\begin{pmatrix}
0 & \Delta_j(t) \\
\Delta^*_j(t) & 0
\end{pmatrix}
\notag\\
&=
-iU
\begin{pmatrix}
0 & \big[\hat{\bm{G}}^<(t)\big]_{2j-1, 2j} \\
\big[\hat{\bm{G}}^<(t)\big]_{2j, 2j-1} & 0
\end{pmatrix},
\label{eq.Delta}
\end{align}
with $\hat{\bm{G}}^<(t) \equiv \hat{\bm{G}}^<(t, t)$ is the equal-time lesser Green's function. In Eqs.~\eqref{eq.self.lead.R} and \eqref{eq.self.lead.<}, the parameter
\begin{equation}
\gamma = \pi \rho |t_\perp|^2
\end{equation}
characterizes the system-reservoir coupling strength. Here, $\rho_\alpha(\omega) = \rho$ is the single-particle density of state in
the $\alpha$ reservoir. We note that, in deriving $\hat{\bm{\Sigma}}^{\rm X}_{\rm lead}$ in Eqs.~\eqref{eq.self.lead.R} and \eqref{eq.self.lead.<}, we have ignored the $\omega$ dependence of $\rho_\alpha(\omega)$ in the $\alpha$ reservoir around the Fermi levels, which is also referred to as the wide-band approximation in the literature~\cite{StefanucciBook}.

\subsection{Dynamics of the superconducting order parameter}

The dynamics of the superconducting order parameter at the $j$th lattice site, $\Delta_j(t>0)$, after switching on the bias voltage $\Delta V(t)$ at $t=0$ is determined by the equation of motion of the equal-time lesser Green's function $\hat{\bm{G}}^<(t)$, which is directly related to the order parameter $\Delta_j(t)$ via Eq.~\eqref{eq.Delta}. Substituting the self-energies in Eqs.~\eqref{eq.self.int.R}-\eqref{eq.self.lead.<} into the Dyson-Keldysh equations~\eqref{eq.Dyson.eq} and \eqref{eq.Keldysh.eq}, we obtain the following equation of motion~\cite{Kawamura2024}:
\begin{align}
i \frac{\partial}{\partial t} \hat{\bm{G}}^<(t) 
&=
\big[\hat{\bm{\m{H}}}_{\rm BdG}(t), \hat{\bm{G}}^<(t)\big]_-  
\notag\\
&\hspace{1cm}
-4i\gamma \hat{\bm{G}}^<(t) -\hat{\bm{\Pi}}(t) -\hat{\bm{\Pi}}^\dagger(t).
\label{eq.QKE}
\end{align}
Here,
\begin{align}
&
\hat{\bm{\m{H}}}_{\rm BdG}(t) = \hat{\bm{\m{H}}}_0(t) -\hat{\bm{\Delta}}(t),\\
&
\hat{\bm{\Pi}}(t) = \int_{-\infty}^\infty dt_1 \hat{\bm{G}}^{\m{R}}(t,t_1) \hat{\bm{\Sigma}}^<_{\rm lead}(t_1,t).
\label{eq.def.Pi}	
\end{align}
For the derivation of Eq.~\eqref{eq.QKE},  we refer to Ref.~\cite{Kawamura2024}. Since $\hat{\bm{\Pi}}(t)$ in Eq.~\eqref{eq.def.Pi} involves the retarded Green's function $\hat{\bm{G}}^{\m{R}}$, Eq.~\eqref{eq.QKE} is solved together with the Dyson equation~\eqref{eq.Dyson.eq}.

The initial conditions $\hat{\bm{G}}^<(t=0)$ and $\hat{\bm{\Pi}}(t=0)$ for Eq.~\eqref{eq.QKE} is obtained from the Dyson-Keldysh equations~\eqref{eq.Dyson.eq} and \eqref{eq.Keldysh.eq}. When the system thermalizes with the reservoirs at $t=0$, the Green's functions $\hat{\bm{G}}^{\rm X}(t,t')$, as well as the self-energies $\hat{\bm{\Sigma}}^{\rm X}(t,t')$, depend only on the relative time $t-t'$. This simplifies the Eqs.~\eqref{eq.Dyson.eq} and \eqref{eq.Keldysh.eq} as
\begin{align}
&
\hat{\bm{G}}^{\m{R}(\m{A})}(\omega)=
\hat{\bm{G}}^{\m{R}(\m{A})}_0(\omega)
+
\hat{\bm{G}}^{\m{R}(\m{A})}_0(\omega)
\hat{\bm{\Sigma}}^{\m{R}(\m{A})}(\omega)
\hat{\bm{G}}^{\m{R}(\m{A})}(\omega)
\label{eq.Dyson.eq.NESS}
,\\	
&
\hat{\bm{G}}^<(\omega)=
\hat{\bm{G}}^{\m{R}}(\omega)
\hat{\bm{\Sigma}}^<(\omega)
\hat{\bm{G}}^{\m{A}}(\omega)
\label{eq.Keldysh.eq.NESS},
\end{align}
where
\begin{equation}
\hat{\bm{F}}^{\rm X}(\omega)
=
\int_{-\infty}^\infty d(t-t')
e^{i\omega(t-t')}
\hat{\bm{F}}^{\rm X}(t-t'),
\end{equation}
with $\hat{\bm{F}}=\hat{\bm{G}}_0, \hat{\bm{G}}, \hat{\bm{\Sigma}}$. In frequency space, the self-energies in Eqs.~\eqref{eq.self.int.R}-\eqref{eq.self.lead.<} at $t=0$ have the forms,
\begin{align}
&
\hat{\bm{\Sigma}}^{\m{R}(\m{A})}_{{\rm int}}(\omega)=
-\hat{\bm{\Delta}}(t=0)
,\\
&
\hat{\bm{\Sigma}}^<_{{\rm int}}(\omega)=0
,\\
&
\hat{\bm{\Sigma}}^{\m{R}(\m{A})}_{{\rm lead}}(\omega)=
-2i\gamma \hat{\bm{1}}
,\\
&
\hat{\bm{\Sigma}}^<_{{\rm lead}}(\omega)=
2i\gamma 
\left[f\left(\omega-\frac{eV_0}{2}\right) +f\left(\omega+\frac{eV_0}{2}\right)\right]
\hat{\bm{1}}.
\end{align}
Substituting these self-energies into Eqs.~\eqref{eq.Dyson.eq.NESS} and \eqref{eq.Keldysh.eq.NESS}, we obtain the dressed Green's function $\hat{\bm{G}}^{\rm X}(\omega)$ at $t=0$. The initial conditions for Eq.~\eqref{eq.QKE} are then given by
\begin{align}
&
\hat{\bm{G}}^<(t=0) = \int_{-\infty}^\infty \frac{d\omega}{2\pi}\hat{\bm{G}}^<(\omega)
,\\
&
\hat{\bm{\Pi}}(t=0) = \int_{-\infty}^\infty \frac{d\omega}{2\pi}\hat{\bm{G}}^{\m{R}}(\omega) \hat{\bm{\Sigma}}^<(\omega).
\end{align}

We note that the direct computation of Eq.~\eqref{eq.QKE} is a numerically demanding task, as Eq.~\eqref{eq.QKE} is an integrodifferential equation that depends on the past information through $\hat{\bm{\Pi}}(t)$ in Eq.~\eqref{eq.def.Pi}. To overcome this computational difficulty, we employ the auxiliary-mode expansion technique~\cite{Croy2009, Croy2011, Croy2012, Popescu2016, Tuovinen2023, Kawamura2024}. This technique, which was originally developed for studying time-dependent electron transport in normal-metal devices~\cite{Croy2009, Croy2011, Croy2012, Popescu2016, Tuovinen2023} and later extended to superconducting junctions~\cite{Kawamura2024}, enables us to transform the integrodifferential equation~\eqref{eq.QKE} into a more tractable set of ordinary differential equations. For more details about this technique and its implementation, we refer to Ref.~\cite{Kawamura2024}.

\subsection{Steady-state soulution}

The superconductor relaxes to a NESS, when a constant bias voltage $V=V_0 +\Delta V$ is imposed when $t>0$. In the following, we explain the theoretical framework to obtain this NESS solution.

\subsubsection{Uniform superconducting steady state}

When the system relaxes into a uniform superconducting steady state ($\Delta_j = \Delta$),  $\hat{\bm{G}}^{\rm X}(t,t')$ and $\hat{\bm{\Sigma}}^{\rm X}(t,t')$ depend only on the relative position $j-k$ and the relative time $t-t'$. This simplifies the Dyson-Keldysh equations~\eqref{eq.Dyson.eq} and \eqref{eq.Keldysh.eq} as
\begin{align}
&
\bm{G}^{\m{R}(\m{A})}_{\bm{k}}(\omega)=
\bm{{G}}^{\m{R}(\m{A})}_{0\bm{k}}(\omega)
+
\bm{{G}}^{\m{R}(\m{A})}_{0\bm{k}}(\omega)
\bm{\Sigma}^{\m{R}(\m{A})}_{\bm{k}}(\omega)
\bm{G}^{\m{R}(\m{A})}_{\bm{k}}(\omega)
\label{eq.Dyson.RA.3}
,\\
&
\bm{G}^<_{\bm{k}}(\omega)=
\bm{G}^{\m{R}}_{\bm{k}}(\omega)
\bm{\Sigma}^<_{\bm{k}}(\omega)
\bm{G}^{\m{A}}_{\bm{k}}(\omega).
\label{eq.Dyson.<.3}
\end{align}
Here, $\bm{G}^{\rm X}_{\bm{k}}(\omega)$ is the $2\times 2$ matrix Nambu Green's function in momentum $\bm{k}=(k_x, k_y)$ space~\cite{Kawamura2022_Jul, Kawamura2023, Kawamura2024}. In Eq.~\eqref{eq.Dyson.RA.3}, the unperturbed Green's function $\bm{{G}}^{\m{R}(\m{A})}_{0\bm{k}}(\omega)$ in momentum space is given by
\begin{equation}
\bm{{G}}^{\m{R}(\m{A})}_{0\bm{k}}(\omega)=
\frac{1}{\omega_\pm -\xi_{\bm{k}} \bm{\tau}_3 +h\bm{\tau}_0},
\label{eq.G0.w.k}
\end{equation}
where $\xi_{\bm{k}}= -2t \big[\cos(k_x) +\cos(k_y)\big] -\mu_{\rm sys}$ is the kinetic energy of an electron (where the lattice
constant is taken to be unity, for simplicity). In frequency and momentum space, the self-energies in Eqs.~\eqref{eq.self.int.R}-\eqref{eq.self.lead.<} under the constant voltage $V(t)=V$ have the forms
\begin{align}
&
{\bm{\Sigma}}^{\m{R}(\m{A})}_{{\rm int},\bm{k}}(\omega)=
-\Delta \bm{\tau}_1
\label{eq.self.R.int.k.w}
,\\
&
{\bm{\Sigma}}^<_{{\rm int},\bm{k}}(\omega)=0
,\\
&
\bm{\Sigma}^{\m{R}(\m{A})}_{{\rm lead}, \bm{k}}(\omega)=
-2i\gamma \bm{\tau}_0
,\\
&
{\bm{\Sigma}}^<_{{\rm lead}, \bm{k}}(\omega)=
4i\gamma f_{\rm neq}(\omega) \bm{\tau}_0.
\label{eq.self.<.NESS}
\end{align}
Here,
\begin{equation}
f_{\rm neq}(\omega) = \frac{1}{2}\left[f\left(\omega-\frac{eV}{2}\right) +f\left(\omega+\frac{eV}{2}\right)\right]
\end{equation}
represents the nonequilibrium distribution function with the two-step structure at $\omega= \pm eV/2$. The superconducting order parameter $\Delta$ in Eq.~\eqref{eq.self.R.int.k.w} is related to the $(1,2)$ component of the lesser Green's function as
\begin{equation}
\Delta = -iU \sum_{\bm{k}} \int_{-\infty}^\infty \frac{d\omega}{2\pi} \big[\bm{G}^<_{\bm{k}}(\omega)\big]_{12}.
\label{eq.Delta.G<12}
\end{equation}
Substituting Eqs.~\eqref{eq.G0.w.k}-\eqref{eq.self.<.NESS} into the Dyson-Keldysh equations~\eqref{eq.Dyson.RA.3} and \eqref{eq.Dyson.<.3}, we have~\cite{Kawamura2023}
\begin{align}
&
\bm{G}^{\m{R}(\m{A})}_{\bm{k}}(\omega)=\sum_{\eta=\pm} \frac{1}{\omega \pm 2i\gamma -\eta E_{\bm{k}}+ h}\bm{\Xi}_{\bm{k}}^\eta
\label{eq.GR.k.w}
,\\
&
\bm{G}^<_{\bm{k}}(\omega)=
\sum_{\eta=\pm} \frac{4i\gamma f_{\rm neq}(\omega)}{[\omega -\eta E_{\bm{k}} +h ]^2 +4\gamma^2}\bm{\Xi}_{\bm{k}}^\eta,
\label{eq.G<.k.w}
\end{align}
where
\begin{equation}
\bm{\Xi}^{\pm}_{\bm{k}} =
\frac{1}{2}\left[ \bm{\tau}_0 \pm \frac{\xi_{\bm{k}}}{E_{\bm{k}}} \bm{\tau}_3 \mp \frac{\Delta}{E_{\bm{k}}} \bm{\tau}_1 \right],
\end{equation}
and $E_{\bm{k}}=\sqrt{\xi_{\bm{k}}^2 +\Delta^2}$ describes the Bogoliubov excitations.
Substituting Eq.~\eqref{eq.G<.k.w} into Eq.~\eqref{eq.Delta.G<12}, we obtain the self-consistent equation with respect to $\Delta$ as
\begin{widetext}
\begin{equation}
\frac{\Delta}{U}= \sum_{\bm{k}} \int_{-\infty}^\infty \frac{d\omega}{2\pi}\frac{4\gamma \Delta \big[\omega +h\big]\big[1 -2f_{\rm neq}(\omega)\big]}{\big[(\omega -E_{\bm{k}}+h)^2 +4\gamma^2\big]\big[(\omega +E_{\bm{k}}+h)^2 +4\gamma^2\big]}.
\label{eq.NESS.gap.eq}
\end{equation}
We self-consistently solve the nonequilibrium gap equation~\eqref{eq.NESS.gap.eq}, to determine $\Delta$ for a given parameter set $(\gamma, V, T_{\rm env}, h)$.
\end{widetext}

\subsubsection{Normal steady state \label{sec.Thouless}}
The boundary between the normal phase ($\Delta =0$) and the superconducting phase can be determined by examining the pole of the retarded particle-particle scattering $T$-matrix $\chi^{\m{R}}(\bm{q},\nu)$~\cite{Kawamura2020, Kawamura_AAPPS, Kawamura2023}. As shown by Kadanoff and Martin (KM)~\cite{Kadanoff1961, KadanoffBook}, the normal state becomes unstable (Cooper instability) and the superconducting transition occurs, when the $T$-matrix $\chi^{\m{R}}(\bm{q},\nu)$ exhibits a pole at $(\bm{q},\nu)=(\bm{Q},0)$.

To apply the KM theory to the present nonequilibrium system, we evaluate the $T$-matrix $\chi^{\m{R}}(\bm{q},\nu)$ within the mean-field (ladder) approximation, which reads~\cite{Kawamura2020, Kawamura_AAPPS}
\begin{equation}
\chi^{\m{R}}(\bm{q},\nu)= \frac{-U}{1+U\chi^{\m{R}}_0(\bm{q},\nu)}.
\label{eq.chi.R}
\end{equation}
Here,
\begin{align}
\chi^{\m{R}}_0(\bm{q},\nu)
&=	
i \sum_{\bm{k}} \int_{-\infty}^\infty \frac{d\omega}{2\pi}
\Big[
\m{G}^{\m{R}}_{\bm{k}+\bm{q}/2, \up}(\omega+\nu)
\m{G}^<_{-\bm{k}+\bm{q}/2, \down}(-\omega)
\notag\\
&\hspace{0.5cm}+
\m{G}^<_{\bm{k}+\bm{q}/2, \up}(\omega+\nu)
\m{G}^{\m{R}}_{-\bm{k}+\bm{q}/2, \down}(-\omega)
\notag\\
&\hspace{0.5cm}+
\m{G}^{\m{R}}_{\bm{k}+\bm{q}/2, \up}(\omega+\nu)
\m{G}^{\m{R}}_{-\bm{k}+\bm{q}/2, \down}(-\omega)
\Big]
\label{eq.chi0}
\end{align}
is the lowest-order pair correlation function. In Eq.~\eqref{eq.chi0}, $\m{G}^{\rm X}_{\bm{k}, \sigma}(\omega)$ is the Green's function in the normal steady state. This Green's function is obtained from Eqs.~\eqref{eq.GR.k.w} and \eqref{eq.G<.k.w} by setting $\Delta=0$ and extracting the diagonal components, which yields
\begin{align}
\m{G}^{\m{R}(\m{A})}_{\bm{k}, \sigma}(\omega) &=
\frac{1}{\omega \pm 2i\gamma -\xi_{\bm{k},\sigma}}
,\\[4pt]
\m{G}^<_{\bm{k}, \sigma}(\omega) &=
\frac{4i\gamma f_{\rm neq}(\omega)}{[\omega -\xi_{\bm{k},\sigma}]^2 +4\gamma^2},
\label{eq.G<.N.k}
\end{align}
where $\xi_{\bm{k}, \sigma} = \xi_{\bm{k}} -\sigma h$. From Eq.~\eqref{eq.chi.R}, the pole of the $T$-matrix is determined from the equation,
\begin{equation}
1 + U \chi_0^{\m{R}}(\bm{q}=\bm{Q}, \nu=0) =0.
\label{eq.KM}
\end{equation}
Substituting Eqs.~\eqref{eq.chi0}-\eqref{eq.G<.N.k} into Eq.~\eqref{eq.KM}, we obtain the KM condition as
\begin{widetext}
\begin{equation}
\frac{1}{U} = \sum_{\bm{k}} \int_{-\infty}^\infty \frac{d\omega}{2\pi} \frac{2\gamma\big[\omega -\xi^{\rm a}_{\bm{k}, \bm{Q}}\big]\left[\tanh\left(\frac{\omega+eV/2}{2T_{\rm env}}\right) +\tanh\left(\frac{\omega-eV/2}{2T_{\rm env}}\right)\right]}{\big[(\omega -\xi_{\bm{k}+\bm{Q}/2, \up})^2 +4\gamma^2\big]\big[(\omega +\xi_{-\bm{k}+\bm{Q}/2, \down})^2 +4\gamma^2\big]}.
\label{eq.Thouless}
\end{equation}
\end{widetext}
Here,
\begin{equation}
\xi^{\rm a}_{\bm{k}, \bm{Q}} = \frac{1}{2}\big[\xi_{\bm{k}+\bm{Q}/2,\up} -\xi_{-\bm{k}+\bm{Q}/2, \down} \big]
\end{equation}
and the momentum $\bm{Q}$ is determined so as to obtain the highest critical temperature $T_{\rm env}^{\rm c}$.  The solution with $\bm{Q}=0$ gives the onset of the BCS-type uniform superconducting steady state. Indeed, Eq.~\eqref{eq.Thouless} with $\bm{Q}=0$ just equals the nonequilibrium gap equation~\eqref{eq.NESS.gap.eq} with $\Delta=0$. On the other hand, the solution with $\bm{Q}\neq 0$ corresponds to the onset of the inhomogeneous FFLO-type superconducting steady state~\cite{Kawamura2020, Kawamura_AAPPS, Kawamura2023}.

\begin{figure*}[t]
\centering
\includegraphics[width=16.5cm]{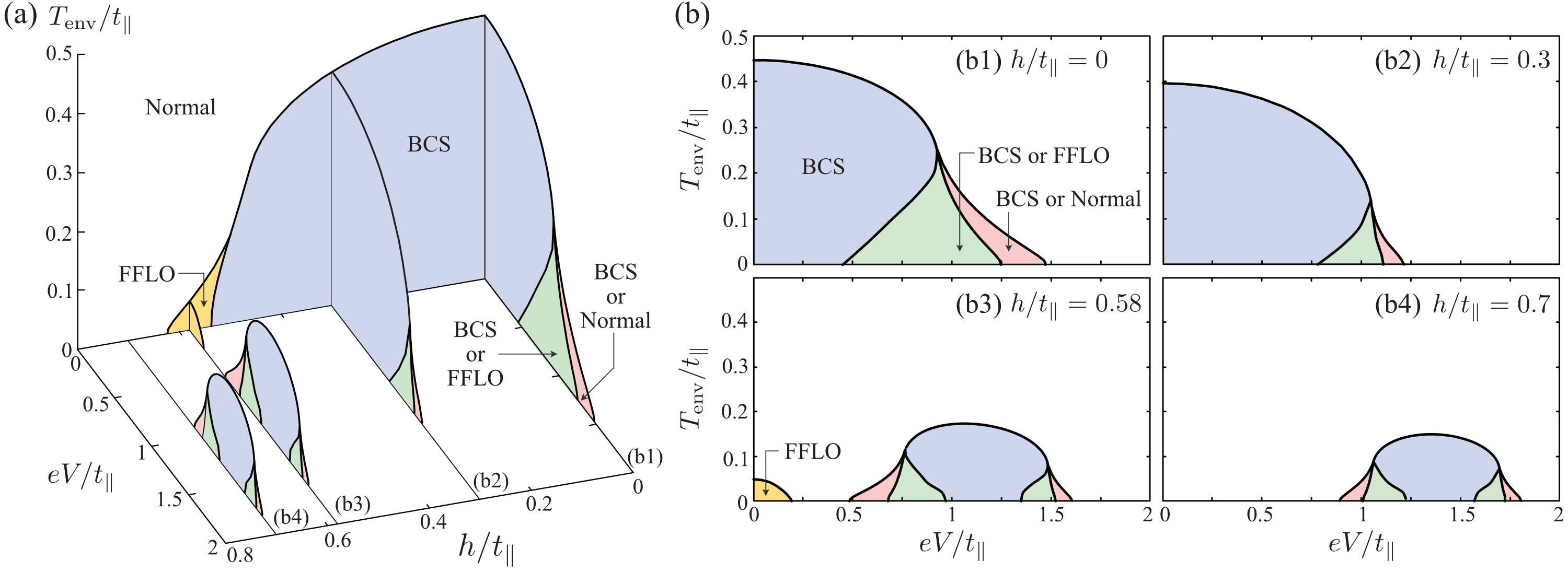} 
\caption{(a) Phase diagram of a superconductor under an external magnetic field $h$ and a bias voltage $V$. BCS (FFLO) is phase of the uniform (non-uniform) superconducting steady state. In the green and red regions, the system exhibits bistability: when entering these regions from the normal phase (white region), the system remains in the normal state in the red region, and transitions into the FFLO state in the green region. On the other hand, when entering from the BCS phase (blue region), the BCS state is maintained in both regions. (b) $V$-$T_{\rm env}$ phase diagram for different values of the external magnetic field $h$.}
\label{fig.PD}
\end{figure*}

\section{Phase diagram of the superconductor under an external magnetic field and bias voltage \label{sec.result}}

We now explore the nonequilibrium phase diagram of the superconductor under an external magnetic field $h$ and a bias voltage $V$. Hereafter, we scale the energy in terms of the hopping amplitude $t_\parallel$. The model parameters are set as $U/t_\parallel=3$, $\mu_{\rm sys}=0$, and $\gamma/t_\parallel=0.01$.

Under a constant magnetic field $h$ and bias voltage $V$, the main system relaxes into a NESS. Figure~\ref{fig.PD} presents the phase diagram summarizing the NESSs of the system. In this phase diagram, the ``BCS" is the uniform superconducting steady state with a constant order parameter ($|\Delta_j| = \Delta$), while the ``FFLO" is the inhomogeneous superconducting steady state characterized by a spatially oscillating order parameter. The system exhibits bistability in the green and red regions, where two distinct steady states are both stable. In these bistable regions, the system relaxes into one of the two steady states depending on how the bias voltage is adjusted, as we will demonstrate later.

At zero bias voltage ($V=0$), the superconductor thermalizes with the normal-metal leads. Consequently, the $h$-$T_{\rm env}$ phase diagram at $V=0$, shown in Fig.~\ref{fig.PD}(a), exhibits the same structure as that of an isolated superconductor under an external magnetic field~\cite{Fulde1964, Larkin1964, Takada1969, Matsuda2007}. In the low-temperature and high-magnetic-field regime (yellow region), the FFLO state stabilizes due to Zeeman-split energy bands induced by the external magnetic field, as illustrated in Fig.~\ref{fig.mechanism}(a). We summarize how to determine the $h$-$T_{\rm env}$ phase diagram at $V=0$ in Appendix~\ref{sec.app}.

Figure~\ref{fig.PD}(b1) shows the $V$-$T_{\rm env}$ phase diagram at $h=0$, which was obtained in Ref.~\cite{Kawamura2024}. In the green region, the FFLO state emerges due to the two-step nonequilibrium distribution function, as illustrated in Fig.~\ref{fig.mechanism}(b).

Figures~\ref{fig.PD}(b2)-(b4) present the $V$-$T_{\rm env}$ phase diagram in the presence of an external magnetic field $h$. When $h/t_\parallel=0.58$, we see in Fig.~\ref{fig.PD}(b3) that the thermal equilibrium FFLO state induced by Zeeman-split energy bands (yellow region) is suppressed when the system is driven out of equilibrium by the applied voltage, leading to a transition into the normal steady state. However, as the voltage is further increased, superconductivity reemerges. When $h/t_\parallel=0.7$, as shown in Fig.~\ref{fig.PD}(b4), the strong external magnetic field completely destroys superconductivity in the thermal equilibrium state ($V=0$). Nevertheless, superconductivity is restored as the bias voltage is applied.

In the following, we consider the case when $h/t_\parallel=0.58$, shown in Fig.~\ref{fig.PD}(b3), as an example to illustrate the procedure for determining the nonequilibrium phase diagram in Fig.~\ref{fig.PD}. We also discuss the mechanism underlying the reentrant phenomenon of the superconducting state as the bias voltage is increased.

\begin{figure*}[t]
\centering
\includegraphics[width=16.5cm]{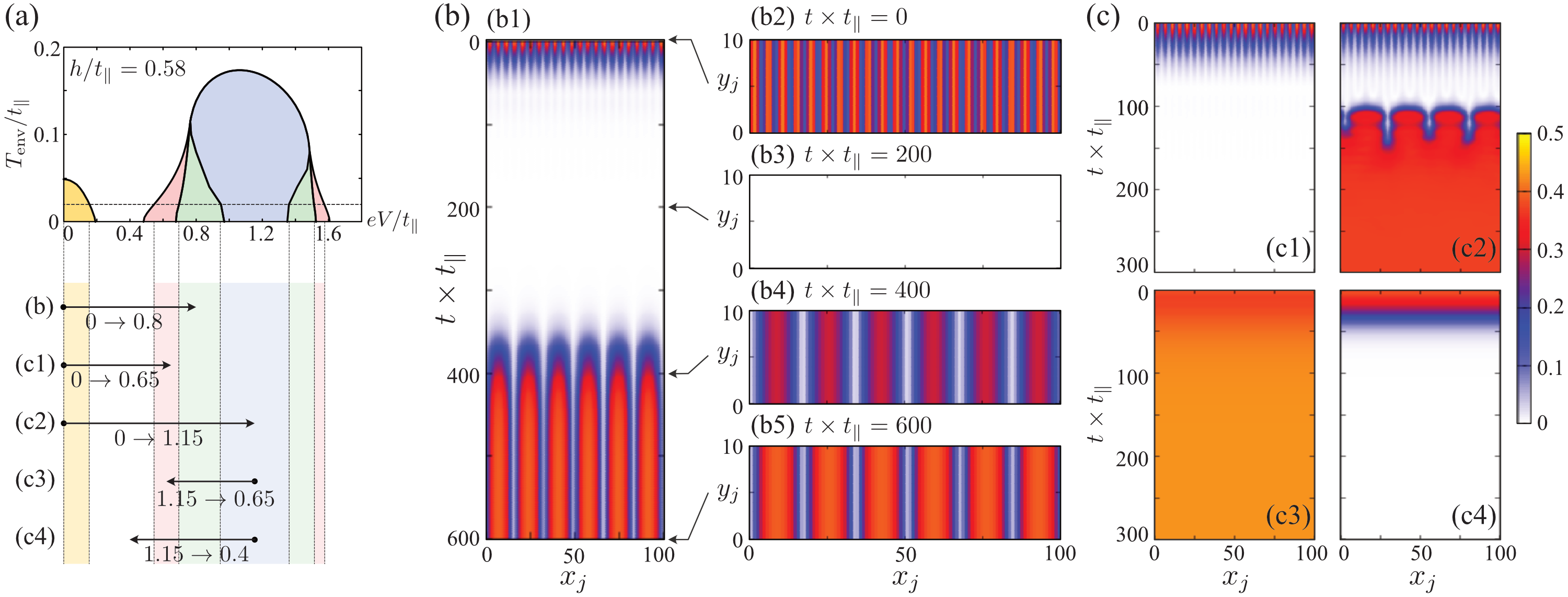}
\caption{Time evolution of the superconducting order parameter following a voltage quench at $t=0$. (a) Schematic diagram illustrating the voltage quench protocol. All simulations are performed at $T_{\rm env}/t_\parallel =0.02$. (b) Dynamics of the order parameter when the voltage is quenched from $e V=0$ (yellow region) to $e V=0.8 t_\parallel$ (green region). (b1) Time evolution of the order parameter amplitude $|\bar{\Delta}_x(t)|$, averaged over the $y$ direction, as defined in Eq.~\eqref{eq.def.OP.x}. (b2)-(b5) Order parameter amplitude $|\Delta_{x,y}(t)|$ in the $x$-$y$ plane at different time steps. (c) Evolution of $|\bar{\Delta}_x(t)|$ for each voltage quench shown in (a). In panels (b) and (c), the amplitude of $|\bar{\Delta}_x(t)|$ and $|\Delta_{x,y}(t)|$ are normalized by $\Delta_0$, which represents the order parameter in the case when the superconductor is isolated from the normal-metal leads and is in the BCS ground state.
}
\label{fig.dynamics} 
\end{figure*}

Figure~\ref{fig.dynamics} shows the dynamics of the superconducting order parameter following the voltage quench at $t=0$. We investigate the spatial variation of the order parameter along the $x$ axis by solving Eq.~\eqref{eq.QKE} on the square lattice with $N = L_x \times L_y = 101 \times 11$ sites. In Figs.~\ref{fig.dynamics}(b) and (c), we show the time evolution of the averaged magnitude of the order parameter  over the $y$ direction, defined by
\begin{equation}
|\bar{\Delta}_x(t)| = \frac{1}{L_y} \sum_{y=1}^{L_y} |\Delta_{x,y}(t)|,
\label{eq.def.OP.x}
\end{equation}
where $(x, y)$ denotes the lattice site coordinates.

Figure~\ref{fig.dynamics}(b) shows the result, when the voltage is quenched from $e V=0$ (yellow region) to $e V=0.8 t_\parallel$ (green region) at $t=0$, as indicated in Fig.~\ref{fig.dynamics}(a). At $t=0$, the main system is initially in the thermal equilibrium FFLO state, as shown in Fig.~\ref{fig.dynamics}(b2). Following the voltage quench, the order parameter amplitude is suppressed and the system transitions into the normal state, as shown in Fig.~\ref{fig.dynamics}(b3). Subsequently, as shown in Fig.~\ref{fig.dynamics}(b4) and (b5), the order parameter amplitude recovers, and the system relaxes into an FFLO state characterized by a spatially oscillating order parameter. In contrast, as shown in Fig.~\ref{fig.dynamics}(c1), when the voltage is quenched from $eV=0$ to $eV=0.65 t_\parallel$ (red region), the superconducting state no longer revives, once the thermal equilibrium FFLO state is destroyed.

\begin{figure}[t]
\centering
\includegraphics[width=7.8cm]{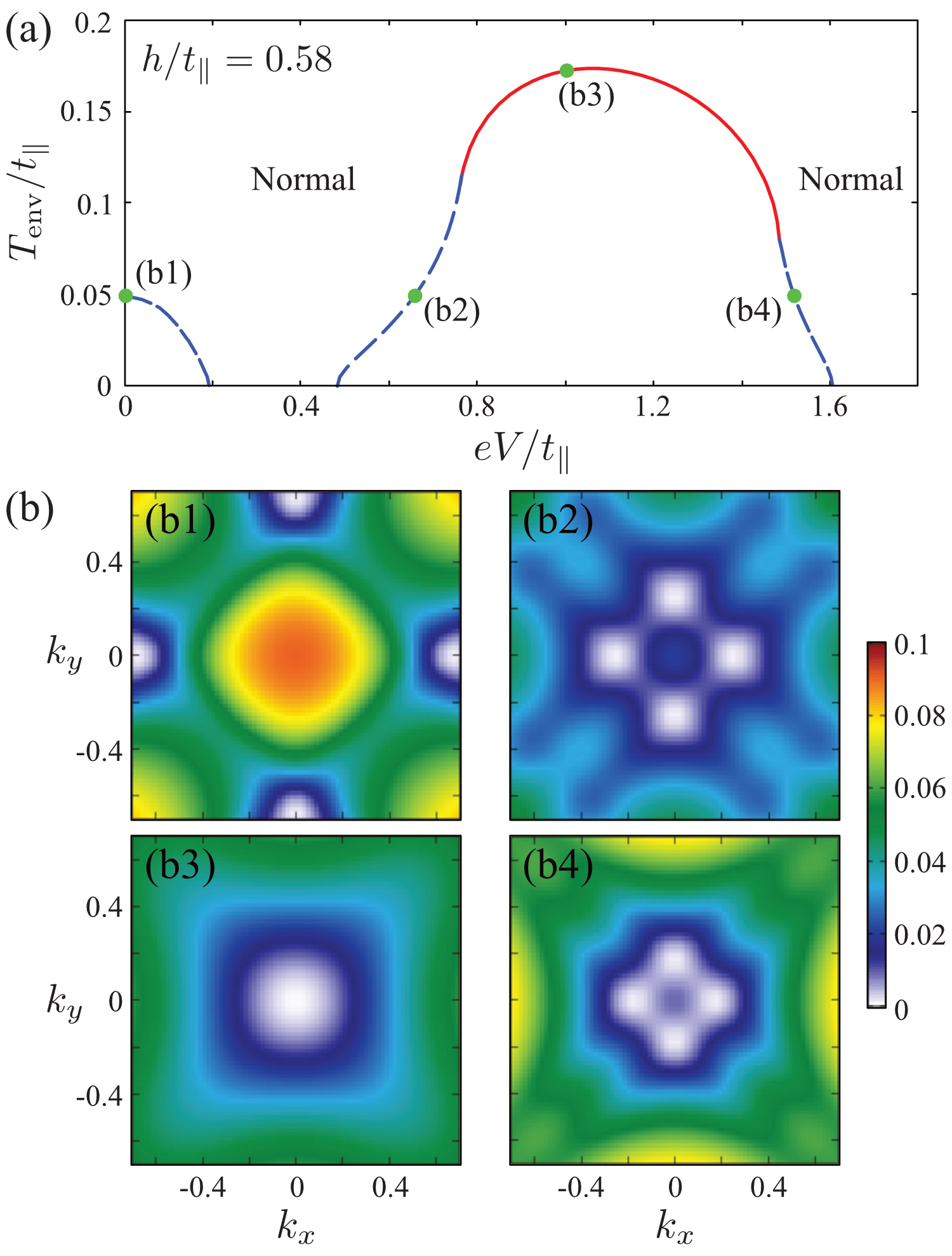}
\caption{(a) Calculated onset of the superconducting steady states for $h/t_\parallel =0.58$. The red solid line (blue dashed line) denotes the onset of the BCS (FFLO) state, corresponding to the boundary between the white and blue regions (the white and yellow regions, as well as the red and green regions) in Fig.~\ref{fig.PD}(b3). (b) The inverse of the $T$-matrix $\big[\chi^{\m{R}}(\bm{q}, \nu=0)/t_\parallel\big]^{-1}$ at points (b1)-(b4) in panel (a). While the $T$-matrix exhibits a pole at $\bm{Q}=0$ at (b3), it has poles at $\bm{Q}\neq 0$ in the other cases.}
\label{fig.Thouless}
\end{figure}

The onset of the FFLO state (the boundaries between the white and the yellow regions, as well as between the red and the green regions) can be determined by solving Eq.~\eqref{eq.Thouless}. As shown in Fig.~\ref{fig.Thouless}(b1), (b2), and (b4), the inverse of the $T$-matrix $\big[\chi^{\m{R}}(\bm{q}, \nu=0)\big]^{-1}$ in Eq.~\eqref{eq.chi.R} vanishes at $\bm{Q}\neq 0$ along the dashed line in Fig.~\ref{fig.Thouless}(a). As explained in Sec.~\ref{sec.Thouless}, this dashed line represents the phase transition from the normal to the FFLO-type inhomogeneous superconducting steady state, corresponding to the boundaries between the white and the yellow regions, as well as between the red and the green regions, in Fig.~\ref{fig.PD}(b3). We note that along the solid line in Fig.~\ref{fig.Thouless}(a), the $T$-matrix has a pole at $\bm{Q}=0$, as seen in Fig.~\ref{fig.Thouless}(b3). Thus, it is identified as the phase transition line from the normal to the BCS-type uniform superconducting steady state, corresponding to the boundary between the white and the blue regions in Fig.~\ref{fig.PD}(b3).

\begin{figure}[t]
\centering
\includegraphics[width=7.7cm]{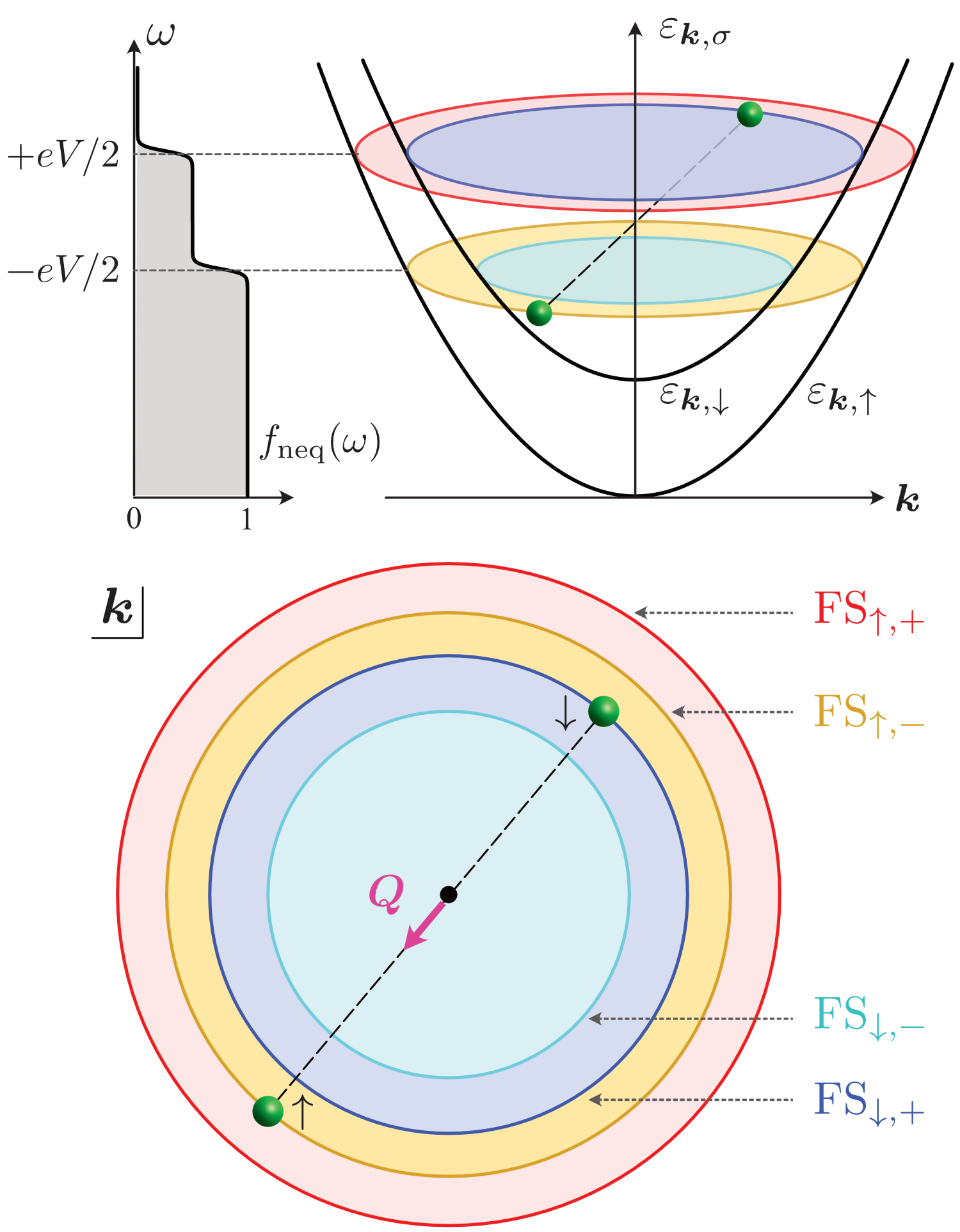}
\caption{Mechanism for the recovery of the FFLO state with increasing bias voltage. The combination of Zeeman-split energy bands ($\ep_{\bm{k},\up} \neq \ep_{\bm{k},\down}$) and the two-step nonequilibrium distribution function $f_{\rm neq}(\omega)$ gives rise to four effective ``Fermi surfaces" (${\rm FS}_{\up,+}$, ${\rm FS}_{\up,-}$, ${\rm FS}_{\down,+}$, and ${\rm FS}_{\down,-}$) with different sizes. The reemergence of the FFLO state is driven by Cooper pairing between ${\rm FS}_{\up, -}$ and ${\rm FS}_{\down, +}$.}
\label{fig.four.FSs}
\end{figure}

The reentrant of the FFLO state seen in Fig.~\ref{fig.dynamics}(b) arises from the combined effects of Zeeman-split energy bands induced by the external magnetic field with the two-step nonequilibrium distribution function caused by the imposed bias voltage. The former splits the spin-$\up$ and -$\down$ Fermi surfaces, which are further split by the latter. As a result, the interplay between the external magnetic field and the bias voltage produces four effective ``Fermi surfaces" (${\rm FS}_{\up,+}$, ${\rm FS}_{\up, -}$, ${\rm FS}_{\down, +}$, and ${\rm FS}_{\down, -}$)  with different sizes~\cite{Kawamura2023}, as illustrated in Fig.~\ref{fig.four.FSs}. The FFLO state that reemerges in the green region in Fig.~\ref{fig.PD}(b3) originates from Cooper pairings between ${\rm FS}_{\up, -}$ and ${\rm FS}_{\down, +}$ among these four effective ``Fermi surfaces", as shown in Fig.~\ref{fig.four.FSs}.

As the voltage is further increased from the situation illustrated in Fig.~\ref{fig.four.FSs}, ${\rm FS}_{\up, -}$ and ${\rm FS}_{\down, +}$ become almost the same size, leading to Cooper pairs with zero center-of-mass momentum between these ``Fermi surfaces". These Cooper pairs induce the BCS-type uniform superconducting steady state (with $\bm{Q}=0$). Indeed, when the voltage is quenched from $eV=0$ (yellow region) to $eV=1.15 t_\parallel$ (blue region), the system relaxes into the uniform BCS state, as shown in Fig.~\ref{fig.dynamics}(c2).

As the voltage is further increased, ${\rm FS}_{\down, +}$ becomes larger than ${\rm FS}_{\up, -}$. This mismatch between the ``Fermi surfaces" once again stabilizes the FFLO state in the green region for $1.35 \lesssim eV/t_\parallel \lesssim 1.5 $, as shown in Fig.~\ref{fig.PD}(b3). When the voltage is decreased from the normal phase ($eV/t_\parallel \gtrsim 1.6$) into the red ($1.5 \lesssim eV/t_\parallel \lesssim 1.6$) and the green ($1.35 \lesssim eV/t_\parallel \lesssim 1.5$) regions, the system remains in the normal steady state in the red region, whereas it transitions into the FFLO state in the green region (although we do not explicitly show the results here).

\begin{figure}[t]
\centering
\includegraphics[width=7.8cm]{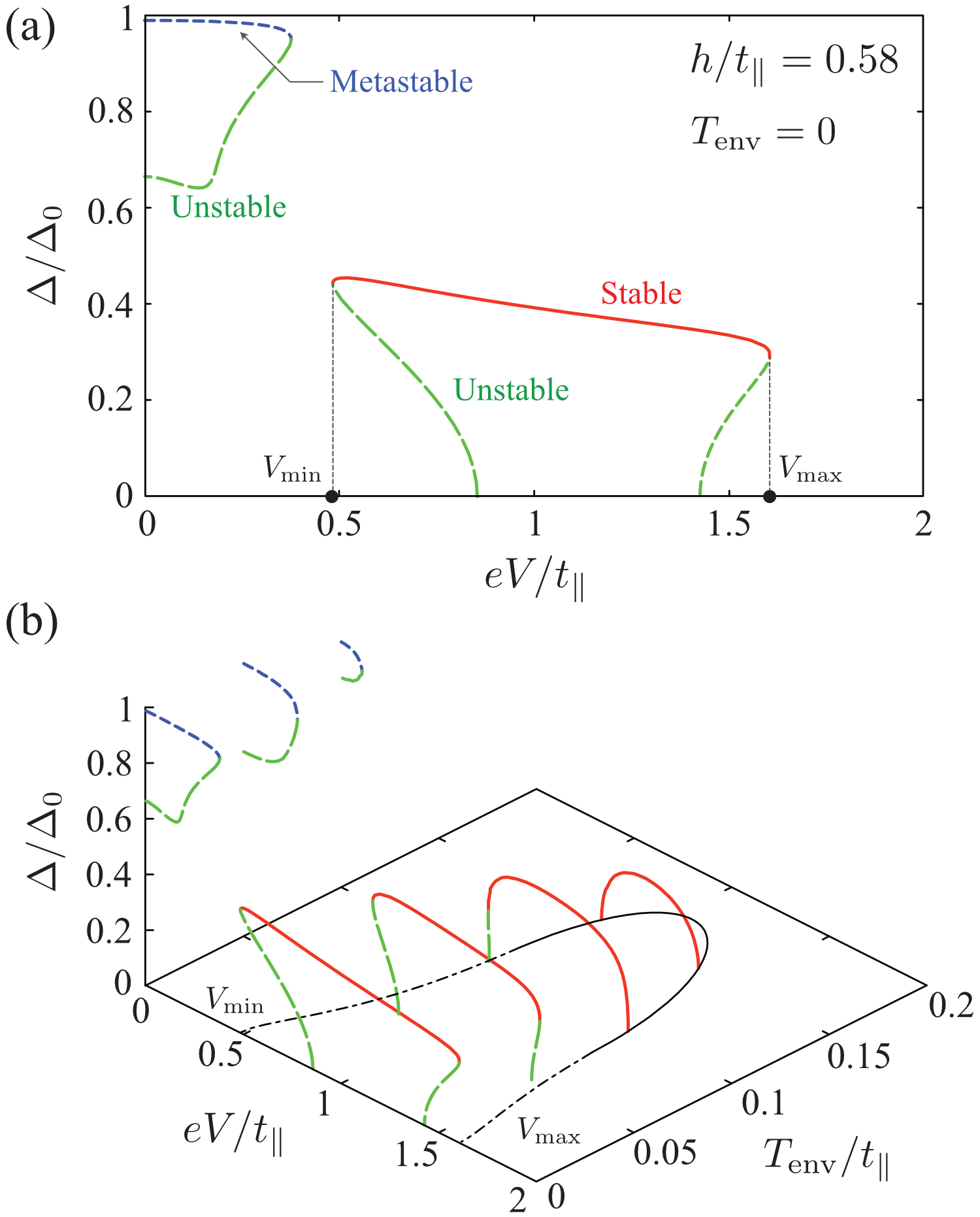}
\caption{Steady-state solutions of the nonequilibrium gap equation~\eqref{eq.NESS.gap.eq} for $h/t_\parallel=0.58$. (a) Solution at $T_{\rm env}=0$. The red solid line represents the stable solution, while the green dashed and blue dotted lines correspond to the unstable and metastable solutions, respectively~\cite{note.meta}. (b) Solution at finite temperature $T_{\rm env}$. The dash-dotted lines ($V_{\rm min}$ and $V_{\rm max}$) in the $V$-$T_{\rm env}$ plane represent the minimum and maximum bias voltages where the stable solution exists, as shown in (a). These dash-dotted lines give the boundaries between the red and the white regions in Fig.~\ref{fig.PD}(b3).
}
\label{fig.gap}
\end{figure}

Figures~\ref{fig.dynamics}(c3) and (c4) show the time evolution of $|\Delta_x(t)|$, when the voltage is decreased from the BCS state at $eV/t_\parallel=1.15$. As shown in Fig.~\ref{fig.dynamics}(c3), the uniform BCS state is maintained in the red region. The BCS state is also maintained in the green region (although we do not explicitly show the result here). On the other hand, we see in Fig.~\ref{fig.dynamics}(c4) that the system transitions into the normal steady state upon entering the white region.

The boundary between the red and the white regions in Fig.~\ref{fig.PD}(b3), where the phase transition from the BCS state to the normal state occurs, is determined from the nonequilibrium gap equation~\eqref{eq.NESS.gap.eq}. Figure~\ref{fig.gap}(a) shows the superconducting order parameter $\Delta$ as a function of the voltage $V$ in the BCS state at $T_{\rm env}=0$, which is obtained by solving Eq.~\eqref{eq.NESS.gap.eq}. The solid line represents the stable solution, corresponding to the BCS state realized in the blue region in Fig~\ref{fig.PD}(b3). In contrast, the dashed and dotted lines indicate unstable and metastable solutions, respectively, which are not realized as a NESS~\cite{note.meta}. The minimum and maximum voltage ($V_{\rm min}$ and $V_{\rm max}$) for the stable solution, marked by the dash-dotted line in the $V$-$T_{\rm env}$ plane of Fig.~\ref{fig.gap}(b), give the boundaries between the red and the white regions in Fig.~\ref{fig.PD}(b3). The solid line in the $V$-$T_{\rm env}$ plane of Fig.~\ref{fig.gap}(b) is identical to the solid line in Fig.~\ref{fig.Thouless}(a), which gives the boundary between the blue and the white regions in Fig.~\ref{fig.PD}(b3).

Summarizing the above results, we obtain the phase diagram shown in Fig.~\ref{fig.PD}. As demonstrated in Fig.~\ref{fig.dynamics}, the steady state realized in the bistable (red and green) regions depends on how the voltage is varied. When entering these regions from the normal phase (white region), the system remains in the normal state in the red region, whereas it transitions into the FFLO-type inhomogeneous superconducting state in the green region. Conversely, when entering from the BCS phase (blue region), the BCS state is preserved in both regions.

At the end of this section, we briefly discuss how the above results are affected by the boundary of the present model. So far, we have focused on the modulation of the superconducting order parameter along the $x$ axis by setting $L_x=101 > L_y=11$, thereby neglecting the possibility of a two-dimensional oscillation of the order parameter, such as $\Delta_{x,y}=\Delta\big[\sin(Qx) +\sin(Qy) \big]$. However, we expect that our results remain unaffected by this choice of the boundary. In thermal equilibrium superconductors under an external magnetic field, the possibility of two-dimensional oscillations of the order parameter was investigated in Ref.~\cite{Wang2006}, which concludes that the unidirectional FFLO state is always energetically favored over such two-dimensional FFLO states. Similarly, the possibility of two-dimensional FFLO states has been examined in a voltage-driven nonequilibrium superconductor in the absence of an external magnetic field~\cite{Kawamura2024}, where it is shown that the two-dimensional FFLO state is a linearly stable but nonlinearly unstable state. Thus, the inhomogeneous superconducting states observed in the yellow and green regions in Fig.~\ref{fig.PD} are expected to be the unidirectional FFLO state, such as those shown in Fig.~\ref{fig.dynamics}(b2) and (b5), rather than two-dimensional FFLO states.

\section{Summary}

In summary, we have explored nonequilibrium properties of a normal metal-superconductor-normal metal (NSN) junction under an external magnetic field. When a bias voltage is applied between the normal-metal leads, the confined superconductor is driven out of equilibrium, resulting in a nonequilibrium quasiparticle distribution function having a two-step structure. Using the nonequilibrium Green’s function technique, we have developed theoretical frameworks to determine the nonequilibrium phase diagram of the superconductor.

The obtained nonequilibrium phase diagram is presented in Fig.~\ref{fig.PD}. The interplay between Zeeman-split energy bands and the two-step nonequilibrium distribution function gives rise to a rich phase structure. Notably, we found that superconductivity destroyed by a strong external magnetic field can be restored by applying the bias voltage. We explained that this reentrant phenomenon originates from the four effective ``Fermi surfaces" that arise from the combination of Zeeman-split energy bands and the two-step structure in the nonequilibrium distribution function.

We also identified several bistable regions in the phase diagram. By explicitly analyzing the time evolution of the superconducting order parameter following a voltage quench, we demonstrated that the system relaxes into different nonequilibrium steady states in these bistable regions. The realized steady state depends on how the bias voltage is tuned.

Our results demonstrate the potential for controlling quantum states of matter by appropriately designing both the energy band structure and the distribution function describing the occupied states. The realization of unconventional ordered phases through such quantum state engineering is one of the most exciting challenges in modern condensed matter physics, and our findings would contribute to the further development of this research field.

\medskip
\textbf{Acknowledgements}\\
We thank S. Sumita and Y. Kato for stimulating discussions. T.K. was supported by MEXT and JSPS KAKENHI Grant-in-Aid for JSPS fellows (Grant No. JP24KJ0055). Y.O. was supported by a Grant-in-Aid for Scientific Research from MEXT and JSPS in Japan (Grant No. JP22K03486). Some of the computations in this work were done using the facilities of the Supercomputer Center, the Institute for Solid State Physics, the University of Tokyo.

\medskip
\textbf{Conﬂict of Interest}\\
The authors declare no conﬂict of interest.

\medskip
\textbf{Data Availability Statement}\\
The data that support the ﬁndings of this study are available from the corresponding author upon reasonable request.

\medskip
\textbf{Keywords}\\
Unconventional superconductivity, Nonequilibrium superconductivity, FFLO

\appendix
\section{$h$-$T_{\rm env}$ phase diagram at $V=0$ \label{sec.app}}

\begin{figure}[t]
\centering
\includegraphics[width=7.8cm]{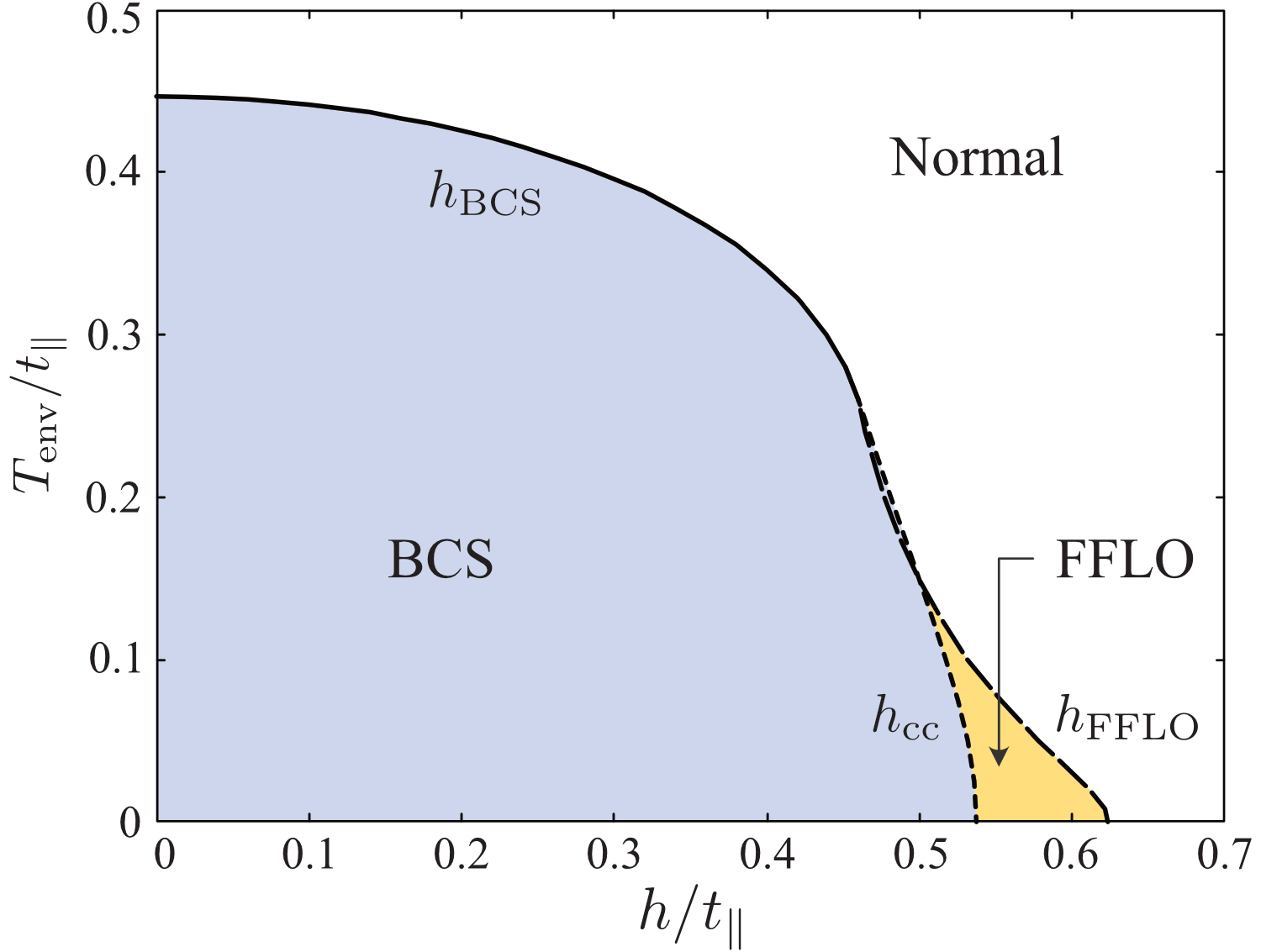}
\caption{$h$-$T_{\rm env}$ phase diagram at $V=0$, shown in Fig.~\ref{fig.PD}(a). The solid line ($h=h_{\rm BCS}$) and the dashed line ($h=h_{\rm FFLO}$) represent the BCS and the FFLO superconducting phase transition, respectively. These lines are determined by solving the KM condition in Eq.~\eqref{eq.Thouless} with $V=0$. The dotted line ($h=h_{\rm cc}$) represents the CC limit, which provides a good approximation for the boundary between the FFLO and the BCS phases.}
\label{fig.V0}
\end{figure}

In this appendix, we explain how to determine the $h$-$T_{\rm env}$ phase diagram at $V=0$, shown in Fig.~\ref{fig.PD}(a). The second-order phase transition from the normal state to the superconducting state is determined by solving the KM condition in Eq.~\eqref{eq.Thouless} with $V=0$. In Fig.~\ref{fig.V0}, the solid line ($h=h_{\rm BCS}$) represents the calculated phase boundary for the BCS state, while the dashed line ($h=h_{\rm FFLO}$) corresponds to the phase boundary for the FFLO state.

We next determine the boundary between the FFLO phase and the BCS phase. For simplicity, we approximate this boundary to the Chandrasekhar-Clogston (CC) limit~\cite{Chandrasekhar1962, Clogston1962}, which corresponds to the first-order phase transition line between the normal phase and the BCS phase.  It is known that the CC limit provides a good approximation for the boundary between the FFLO and the BCS phases~\cite{Takada1969}. To determine the CC limit, we evaluate the thermodynamic potential $\Omega(\mu_{\rm sys}, T_{\rm env}, h, \Delta)$ in the presence of system-reservoir couplings. (We note that, in the absence of the bias voltage, the main system is in the thermal equilibrium state at temperature $T_{\rm env}$ and chemical potential $\mu_{\rm sys}$.) The thermodynamic potential $\Omega$ relates to the filling fraction $n=n_\up+n_\down$ in the main system through
\begin{equation}
n(\mu_{\rm sys}, T_{\rm env}, h, \Delta) = -\left(\frac{\partial \Omega}{\partial \mu_{\rm sys}}\right)_{T_{\rm env}, h, \Delta},
\end{equation}
which leads to
\begin{equation}
\Omega(\mu_{\rm sys}, T_{\rm env}, h, \Delta) =
\frac{\Delta^2}{U} -\int_{-\infty}^{\mu_{\rm sys}} n(x, T_{\rm env}, h, \Delta) dx.
\label{eq.Omega}
\end{equation}
Here, we have used $\Omega(\mu_{\rm sys}=-\infty, T_{\rm env}, h, \Delta)= \Delta^2/U$. The filling fractions $n_{\sigma=\up, \down}$ are obtained from the diagonal components of the lesser Green's function in Eq.~\eqref{eq.G<.k.w}, which yields
\begin{align}
n_\up &=
-i \sum_{\bm{k}} \int_{-\infty}^\infty \frac{d\omega}{2\pi} \big[\bm{G}^<_{\bm{k}}(\omega)\big]_{11}
\notag\\
&=
4\gamma \sum_{\bm{k}} \int_{-\infty}^\infty \frac{d\omega}{2\pi}
\Bigg[
\frac{u^2_{\bm{k}}}{\big[\omega -E_{\bm{k}} +h\big]^2 +4\gamma^2}
\notag\\
&\hspace{2.5cm}
+
\frac{v^2_{\bm{k}}}{\big[\omega +E_{\bm{k}} +h\big]^2 +4\gamma^2}
\Bigg]
f(\omega)
,\\[6pt]
n_\down &=
1 +i \sum_{\bm{k}} \int_{-\infty}^\infty \frac{d\omega}{2\pi} \big[\bm{G}^<_{\bm{k}}(\omega)\big]_{22}
\notag\\
&=	
1-4\gamma\sum_{\bm{k}} \int_{-\infty}^\infty \frac{d\omega}{2\pi}
\Bigg[
\frac{v^2_{\bm{k}}}{\big[\omega -E_{\bm{k}} +h\big]^2 +4\gamma^2}
\notag\\
&\hspace{2.5cm}
+
\frac{u^2_{\bm{k}}}{\big[\omega +E_{\bm{k}} +h\big]^2 +4\gamma^2}
\Bigg]
f(\omega),
\label{eq.n.down}
\end{align}
where
\begin{equation}
u^2_{\bm{k}} = 1-v^2_{\bm{k}} = \frac{1}{2}\left[1 +\frac{\xi_{\bm{k}}}{E_{\bm{k}}}\right].
\end{equation}
We compute the thermodynamic potential $\Omega$ using Eqs.~\eqref{eq.Omega}-\eqref{eq.n.down}. The CC limit $h=h_{\rm cc}$ is then determined from the condition,
\begin{equation}
\Omega(\mu_{\rm sys}, T_{\rm env}, h_{\rm cc}, \Delta=0)
=
\Omega(\mu_{\rm sys}, T_{\rm env}, h_{\rm cc}, \Delta_0>0)
,
\end{equation}
where $\Delta_0$ satisfies the gap equation~\eqref{eq.NESS.gap.eq} with $V=0$. The dotted line in Fig.~\ref{fig.V0} shows the calculated CC limit.

\bibliography{NSN_h_ref}

\begin{thebibliography}{75}
\bibitem [{not()}]{note.meta}%
  \BibitemOpen
  \href@noop {} {}\bibinfo {note} {In the unstable states, a small deviation
  $\delta \Delta (t)$ around the steady-state solution $\Delta_0$ grows over
  time. In contrast, in the metastable state, $\delta \Delta (t)$ decays over
  time, indicating local stability. However, since more stable steady states
  always exist at any bias voltage, the metastable state is never realized as a
  true steady state.}\BibitemShut {Stop}%
\end{thebibliography}
\end{document}